\documentclass[prb,onecolumn,12pt]{revtex4-1}
\usepackage{epsfig}
\usepackage{amssymb}
\usepackage{threeparttable}
\usepackage{graphicx}
\usepackage{natbib}
\usepackage{longtable}
\usepackage{txfonts}
\usepackage{color}

\begin{document}
\title{First-principle calculations of phononic, electronic and optical properties of monolayer arsenene and antimonene allotropes}
\author{Yuanfeng Xu$^1$, Bo Peng$^1$, Hao Zhang$^{1,*}$, Hezhu Shao$^{2}$, Rongjun Zhang$^1$, Hongliang Lu$^3$, David Wei Zhang$^3$ and Heyuan Zhu$^1$}
\affiliation{$^1$Shanghai Ultra-precision Optical Manufacturing Engineering Research Center and Key Laboratory of Micro and Nano Photonic Structures (Ministry of Education), Department of Optical Science and Engineering, Fudan University, Shanghai 200433, China\\
$^2$Ningbo Institute of Materials Technology and Engineering, Chinese Academy of Sciences, Ningbo 315201, China\\
$^3$State Key Laboratory of ASIC and System, Institute of Advanced Nanodevices, School of Microelectronics Fudan University, Shanghai 200433, China}

\begin{abstract}
Recently a stable monolayer of antimony in buckled honeycomb structure called antimonene was successfully grown on 3D topological insulator Bi$_2$Te$_3$ and Sb$_2$Te$_3$, which displays semiconducting properties. By first principle calculations, we systematically investigate the phononic, electronic and optical properties of $\alpha-$ and $\beta-$ allotropes of monolayer arsenene/antimonene. We investigate the dynamical stabilities of these four materials by considering the phonon dispersions. The obtained electronic structures reveal the direct band gap of monolayer $\alpha-$As/Sb and indirect band gap of $\beta-$As/Sb. Significant absorption is observed in $\alpha-$Sb, which can be used as a broad saturable absorber.
\end{abstract}
\maketitle

\section{INTRODUCTION}

Recently two-dimensional materials have been intensely investigated due to the extraordinary properties which can be widely used in the next-generation nanoelectronic and optoelectronic applications\cite{doi:10.1021/am501022x,Aktuerk2015,C5CP00563A}. Graphene is a honeycomb monolayer of carbon atoms, which possesses superior carrier mobility\cite{PhysRevX.4.021029}, but with a small on-off ratio due to the zero band gap, which limits the applications in electronic devices. Black phosphorene\cite{Liu2014, doi:10.1021/nl501658d,Li2014}, a monolayer of phosphorus atoms with puckered structure, induces great attention recently due to the versatile properties such as direct semiconducting property with a moderate band gap\cite{hong2014polarized}, a higher hole mobility comparable to graphene\cite{Qiao2014}, high carrier mobility\cite{chen2016nonequilibrium}, strongly anisotropic transport\cite{Liu2014} and etc. 

The extraordinary properties and the potential applications in the fields of electronics and optoelectronics of black phosphorene have aroused great research interest on other group-V elemental monolayers\cite{PhysRevB.92.125420,SMLL201570033,zheng2016,PhysRevB.91.085423,wang2015electronic}, considering the chemical similarity of elements belonging to the same group in the periodic table. Monolayers of arsenic and antimony with different phases were thus proposed and under investigation by expriments\cite{Wang2015a,PhysRevB.91.085423,Aktuerk2015,Wang2015}. The orthorhombic ($\alpha-$) arsenene (monolayer of arsenic) was predicted to be a semiconducting material with a direct band gap of approximate 1 eV. The orthorhombic arsenene ($\alpha-$As) possesses high carrier mobility\cite{zheng2016, Wang2015} and strong anisotropy\cite{zheng2016}, which make it promising for device applications in the semiconductor industry and solar cells\cite{zheng2016,PhysRevB.91.085423}. Monolayer antimonene including $\alpha-$ and $\beta-$ phases are semiconductors with band gaps suitable for their use in 2D electronics\cite{Aktuerk2015}. The $\beta-$Sb has nearly isotropic mechanical properties, whereas $\alpha-$Sb shows strongly anisotropic charecteristic\cite{Wang2015}.

Experimentally, $\beta-$antimonene(monolayer of antimony) has been successfully realized on the substrates of 3D topological insulator Bi$_2$Te$_3$ and Sb$_2$Te$_3$ respectively\cite{:/content/aip/journal/jap/119/1/10.1063/1.4939281}, and the stability of monolayer antimonene allotropes including $\alpha-$, $\beta-$, $\gamma-$, and $\delta-$phase were examined by phonon dispersion calculations\cite{Wang2015}, which shown that only the  monolayer $\alpha-$ and $\beta-$antimonene ($\alpha/\beta-$Sb) are stable as free-standing monolayers. However, to our knowledge, monolayer arsenene has not been experimentally realized up to now\cite{PhysRevB.91.085423}. Considering those studies of monolayer $\alpha-$ and $\beta-$As/Sb, it lacks systematically investigations of the electronic and optical properties of the four materials with two different structures.

In this work, the density functional theory based first-principles calculations are used to investigate the phonon, electronic and optical properties of   four two-dimensional materials, i.e. $\alpha/\beta-$As and $\alpha/\beta-$Sb. We first optimize the structures and investigate the dynamical stabilities of the optimized structures. Then we calculate the electronic structures of these four materials and show that the value of the band gap of $\alpha-$As/Sb is smaller than that of the $\beta$ counterpart by comparison. Finally we calculate the optical properties of these four materials, and find that  $\alpha-$As/Sb possess strong in-plane anisotropy due to the structure anisotropy, and abundant optical properties. The underlying mechanism of the optical properties is further discussed as well.

\section{Method and computational details}

The calculations are performed using the Vienna \textit{ab-initio} simulation package (VASP) based on density functional theory \cite{Kresse1996}. The calculation is carried out by using the projector-augmented-wave pseudo potential method with a plane wave basis set with a kinetic energy cutoff of 500 eV. The exchange-correlation energy is described by the generalized gradient approximation (GGA) using the Perdew-Burke-Ernzerhof (PBE) functional \cite{Perdew1996}. Since the PBE functional always underestimates the band gap of semiconductors, part of the calculations is also performed using the Heyd-Scuseria-Ernzerhof (HSE06) hybrid functional which is constructed by mixing the PBE and Hartree-Fock (HF) fuctionals. HSE06 improves the precision of band gap by reducing the localization and delocalization errors of PBE and HF functions. Here the screening length of HSE06 is 0.2$\AA^{-1}$ and the mixing ratio of the HF exchange potential is 0.25.

The Brillouin zone integration is performed with a 15$\times $15$\times$1 k mesh for geometry optimization and self-consistent electronic structure calculations\cite{Ma2015}. When optimizing atomic positions under this k mesh, the energy convergence value between two consecutive steps are chosen as 10$^{-6}$ eV and the maximum Hellmann-Feynman force acting on each atom is less than 10$^{-3}$ eV/\AA .  

For $\alpha-$ and $\beta-$As/Sb, we use periodic boundary conditions along the three dimensions, and the vacuum space is around 28$\AA$ and 18$\AA$ along the $z$ direction, respectively, which is large enough to avoid the artifical interaction between atom layers. Furthermore, in the ground state, the configuration of extra-nuclear electron of As is 4s$^{2}$4p$^{3}$ and Sb is 5s$^{2}$5p$^{3}$, respectively.

The phonon dispersion is calculated from the harmonic IFCs using the PHONOPY code\cite{Togo2008}. A 5$\times$5$\times$1 supercell with 5$\times$5$\times$1 k mesh is used for $\alpha-$ and $\beta-$As/Sb to ensure the functional convergence. We also calculate the optical properties of monolayer $\alpha-$ and $\beta-$As/Sb using PBE method on a grid of 21$\times $21$\times$1 k-points.

\section{Results and discussion}

\subsection{Optimized structures of $\alpha-$ and $\beta-$As/Sb }

\begin{figure}
\centering
\includegraphics[width=0.8\linewidth]{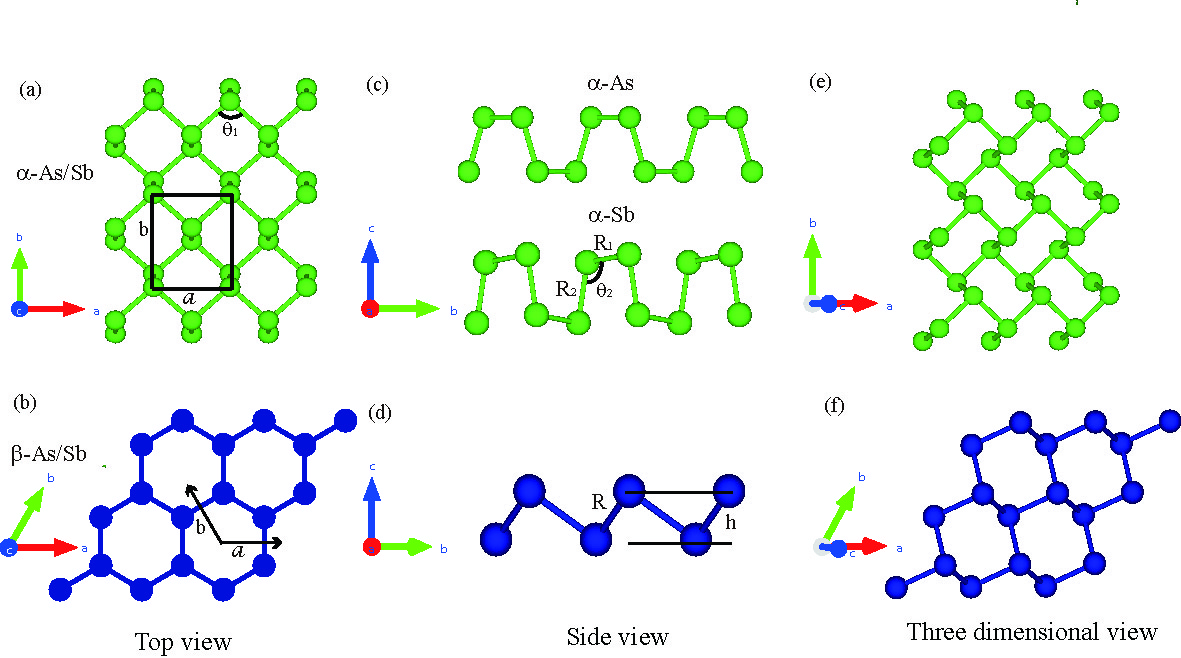}
\caption{(a-b) Top views, (c-d) side views, (e-f) three-dimensional views of the atomic structure of arsenene and antimonene, respectively. }
\label{structure} 
\end{figure}

Fig.~\ref{structure} shows the fully relaxed structures of monolayer $\alpha-$ and $\beta-$As/Sb. The structures of $\alpha-$As and $\alpha-$Sb are puckered with the space group $Pmna$ and $Pmn2$ respectively, and consist of four atoms in a unit cell, as shown in Fig. 1(a). From the side view in Fig.~\ref{structure}(c), both $\alpha-$As and $\alpha-$Sb consist of two atomic sublayers, but for $\alpha-$As all the atoms of a sublayer are located in the same plane. The honeycomb structure of monolayer $\beta-$As/Sb is buckled as well with the space group of $P\overline{3}m1$ and consists of two atoms in each unit cell, as shown in Fig.~\ref{structure}(b). The side view of $\beta-$As/Sb in Fig.~\ref{structure}(d) also shows that two atomic sublayers exist in $\beta-$As/Sb, similar to $\alpha-$As/Sb.

\begin{table}
\centering
\caption{Lattice parameter, bond characteristics and cohesive energy of $\alpha-$As/Sb.}
\begin{tabular}{cccccccccc}
\hline
   & $a$ & $b$ & $R_{1}$ & $R_{2}$ & $\theta_{1}$  & $E_{c}$ & $\theta_{2}$ & $E_{g-\mathrm{PBE}}$/$E_{g-\mathrm{SOC}}$\\
   & ($\AA$) & ($\AA$) & ($\AA$) & ($\AA$) & ($^{\circ}$) & ($^{\circ}$)  & (eV/atom) & (eV)\\
\hline
$\alpha $-As & 3.68(3.71\cite{zheng2016}) & 4.77(4.67\cite{zheng2016}) & 2.51 & 2.50 & 94.50 & 100.66  & -4.62 & 0.91/0.90\\
$\alpha $-Sb & 4.36(4.28\cite{Aktuerk2015}) & 4.73(4.74\cite{Aktuerk2015}) & 2.95 & 2.87 & 95.3 & 102.8  & -4.04 & 0.22/0.21\\
\hline
\end{tabular}
\label{tabel-1}
\end{table}

\begin{table}
\centering
\caption{Lattice parameter, bond characteristics, buckling height and cohesive energy of $\beta-$As/Sb.}
\begin{tabular}{ccccccccc}
\hline
   & $a$  & $R$ & $\theta$ & $h$ & $E_{c}$& $E_{g-\mathrm{PBE}}$/$E_{g-\mathrm{SOC}}$/$E_{g-\mathrm{HSE}}$ \\
   & ($\AA$) & ($\AA$) & ($^{\circ}$) & ($\AA$)   & (eV/atom) & (eV) \\
\hline
$\beta $-As & 3.61(3.607\cite{PhysRevB.91.085423}) &  2.51  & 91.95 & 1.40(1.35\cite{:/content/aip/journal/apl/107/2/10.1063/1.4926761})  & -4.65 & 1.59/1.47/2.11\\
$\beta $-Sb & 4.12(4.12\cite{Wang2015}) & 2.89 & 90.84 & 1.65(1.66\cite{wang2015electronic})   & -4.04 & 1.26/0.99/1.70\\
\hline
\end{tabular}
\label{tabel-2}
\end{table}

Table~\ref{tabel-1} shows the optimized lattice parameters, bond characteristics and cohesive energies of $\alpha-$As/Sb. For $\alpha-$As, the calculated puckered angle is 100.66$^{\circ}$, which is slightly smaller than that of phosphorene (103.69$^{\circ}$)\cite{2053-1583-1-2-025001}. The optimized lattice parameters are $a$ = 3.68\AA\ and $b$ = 4.77$\AA$, which are in good agreement with previous theoretical work\cite{zheng2016}. The calculated bond lengths are 2.50$\AA$ and 2.51$\AA$ with bond angles $\theta_1$ = 94.50$^{\circ}$ and $\theta_2$ = 100.66$^{\circ}$ respectively. For $\alpha-$Sb, the optimized lattice constants are $a$ = 4.36$\AA$ and $b$ = 4.73$\AA$, which are in good agreement with previous theoretical work\cite{Aktuerk2015} as well. Since $a\neq b$, the crystal structure of $\alpha-$As/Sb is anisotropic along $a$ and $b$ directions.

However, for $\beta-$As, the crystal structure is isotropic along the $a$ and $b$ directions, i.e. $a=b$. The optimized lattice parameter $a$ and the buckling height $h$ between the two sublayers are $a=3.61\AA$ and $h=1.40\AA$ respectively, which are in good agreement with previous theoretical results\cite{PhysRevB.91.085423,:/content/aip/journal/apl/107/2/10.1063/1.4926761}. The calculated results lattice parameters and bond characteristics of $\beta-$Sb are shown in Table~\ref{tabel-2}.

\subsection{Phonon dispersions and stabilities of $\alpha-$ and $\beta-$As/Sb }

\begin{figure}
\centering
\includegraphics[width=0.8\linewidth]{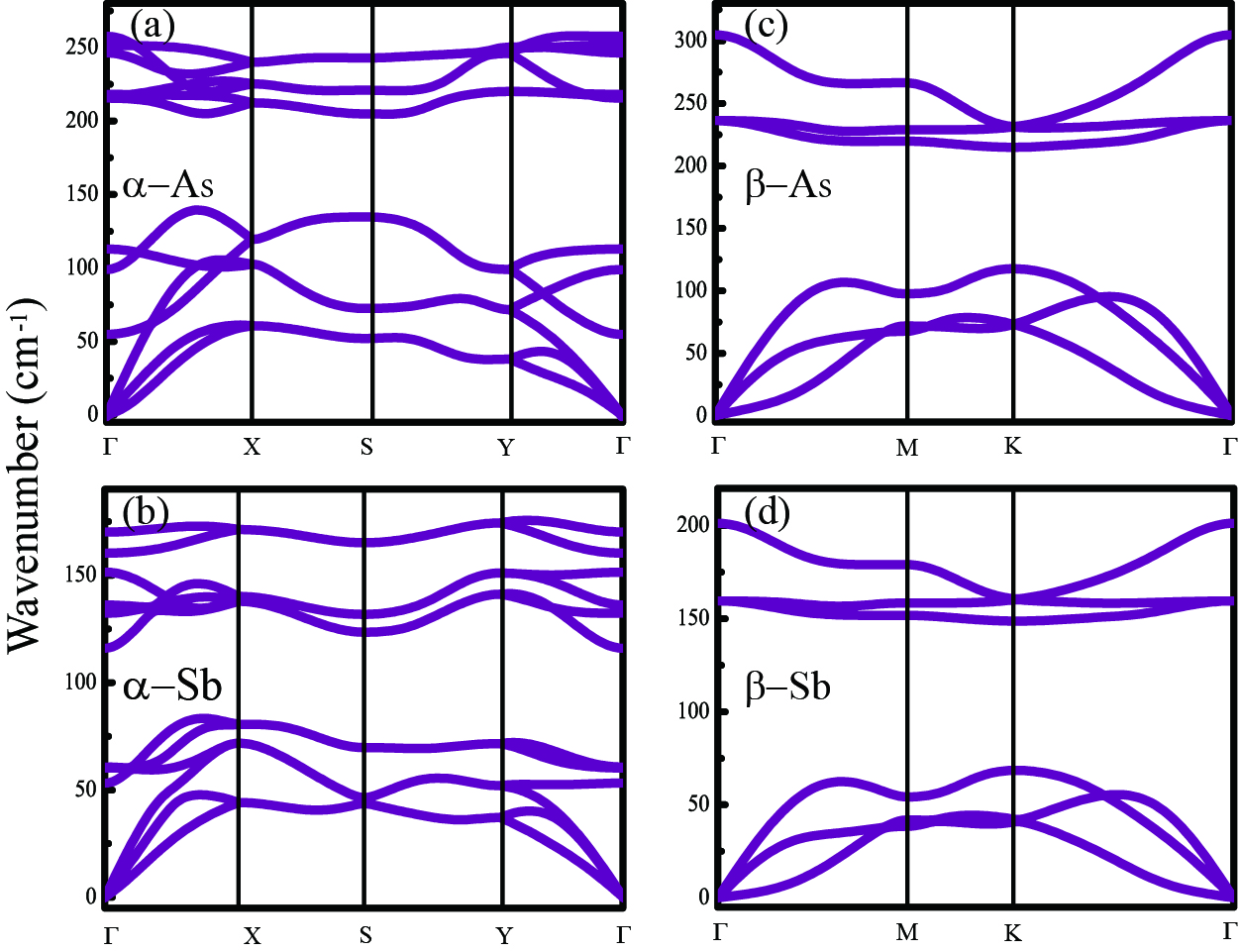}
\caption{Phonon band dispersion of free-standing monolayer (a)$\alpha-$As, (b)$\alpha-$Sb, (c)$\beta-$As and (b)$\beta-$Sb, respectively.}
\label{phonon} 
\end{figure}

For the experimental realization of these newly proposed 2D materials, the consideration of the stability is required. The necessary condition for the structural stabilities of $\alpha-$ and $\beta-$As/Sb against low-frequency acoustic vibrations, which will induce long-wavelength transverse/longitudinal displacements in different directions of the BZ, is that all the phonon frequencies are real. The calculated phonon dispersions of $\alpha-$ and $\beta-$As/Sb are shown in Fig.~\ref{phonon}.

Obviously, there is no imaginary frequency in the obtained phonon dispersion curves for both $\alpha-$As/Sb and $\beta-$As/Sb, revealing the dynamical stability of those four materials as free-standing monolayers. The calculated phonon band gap is about 66 cm$^{-1}$ ($\alpha-$As) , 33 cm$^{-1}$ ($\alpha-$Sb), 97 cm$^{-1}$ ($\beta-$As) and 80 cm$^{-1}$ ($\beta-$Sb), respectively. Since the phonon band gap of $\alpha/\beta-$As is larger than that of $\alpha/\beta-$Sb, more anharmonic three-phonon scatterings among the acoustic phonons and optical phonons in $\alpha/\beta-$As are suppressed compared to $\alpha/\beta-$Sb, which means that the corresponding contribution from the anharmonic three-phonon scatterings to the thermal conductivity in $\alpha/\beta-$As is less important than that in $\alpha/\beta-$Sb. 

In addition, the stability of monolayer $\alpha-$ and $\beta-$As/Sb can be further examined by the cohesive energy calculations. As listed in Table 1 and 2, the cohesive energies of $\alpha$/$\beta-$As are smaller than the corresponding cohesive energies of $\alpha$/$\beta-$Sb, indicating a stronger As-As bond compared to the Sb-Sb bond which is in agreement with the ELF results.

\subsection{Electronic properties of $\alpha-$ and $\beta-$As/Sb }

\begin{figure}
\centering
\includegraphics[width=1.0\linewidth]{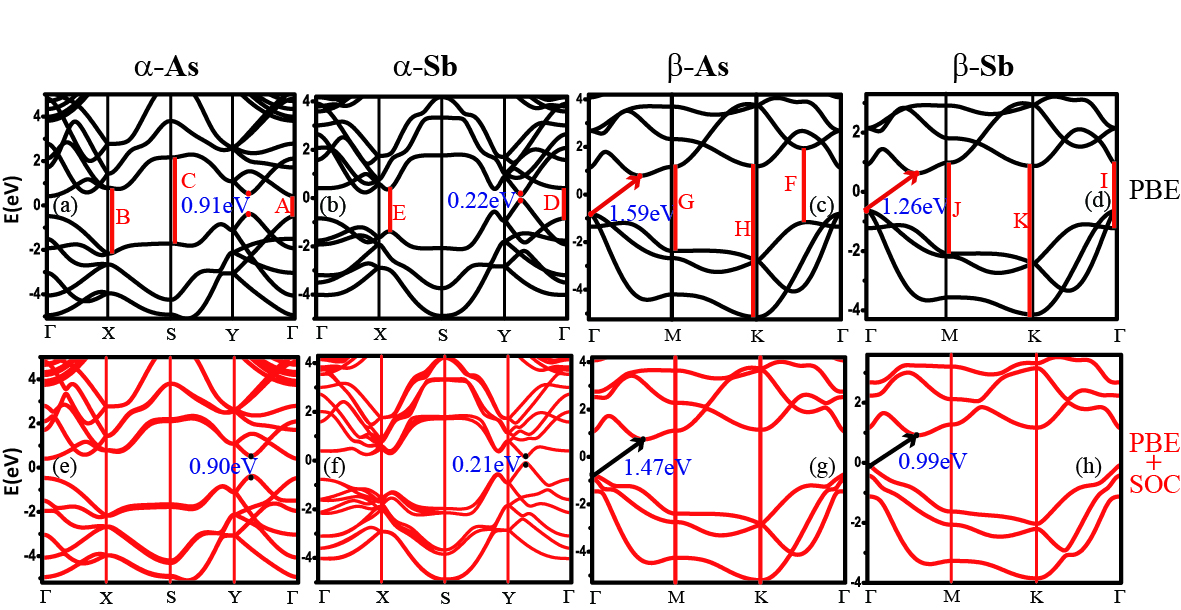}
\caption{Electronic band structures of $\alpha-$ and $\beta-$As/Sb along $\Gamma$-X-S-Y-$\Gamma$ and $\Gamma$-M-K-$\Gamma$ respectively.}
\label{band} 
\end{figure}

The calculated electronic band structures of monolayer $\alpha-$As/Sb and $\beta-$As/Sb along the high-symmetry directions of Brillouin zone (BZ) are shown in Fig.~\ref{band}. Within the PBE functional method, $\alpha-$As is a semiconductor with a direct band gap of 0.91 eV, shown in Fig.~\ref{band}(a), and the minimum of the conduction band (CBM) and maximum of the valance band (VBM) locate within $\Gamma-Y$. The band gap decreases trivially when the spin-orbit coupling (SOC) is involved, as shown in Fig.~\ref{band}(e). Similar calculations are performed on $\alpha-$Sb, shown in Fig.~\ref{band}(b,f). The band gap of $\alpha-$Sb by PBE is 0.22 eV, much smaller than that of $\alpha-$As, which may come from the lower bond energy between Sb atoms than As-As bond. The involved SOC effect gives the band gap of 0.21 eV, as shown in Fig.~\ref{band}(f), and lifts the degeneracy of valence and conduction bands around the Fermi level.

In Fig.~\ref{band}(c) and (d), we calculate the band structures of monolayer $\beta-$As/Sb, and both of which show an indirect band gap between CBM and VBM locating within $\Gamma-M$. The VBM locates at the $\Gamma$ point. For $\beta-$As, the band gap calculated within the PBE functional (Fig.~\ref{band}(c)) is 1.59 eV, reduced to 1.47 eV after SOC effect (Fig.~\ref{band}(g)) involved, and increases to 2.11 eV by considering the HSE06 (Fig.~\ref{band}(k)) correction. Similar calculations of $\beta-$Sb show that the band gap of PBE/PBE+SOC/HSE06 is 1.59/1.47/1.70 eV respectively, as shown in Figs.~\ref{band}(d,h,l).

\begin{figure}
\centering
\includegraphics[width=0.9\linewidth]{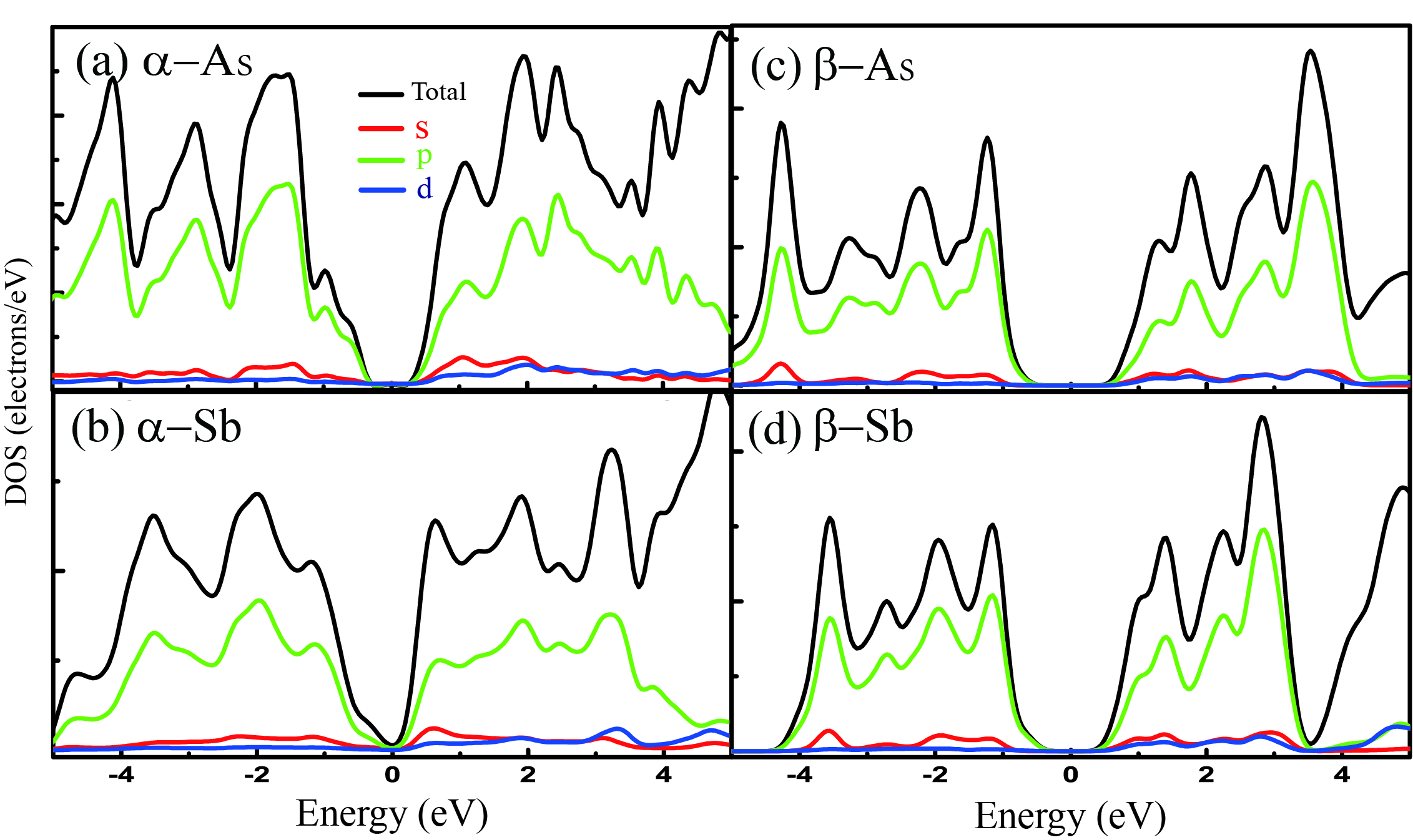}
\caption{Total density of states and partial density of states for monolayer (a) $\alpha-$As, (b) $\alpha-$Sb, (c) $\beta-$As and (d) $\beta-$Sb in the energy range from -5 to 7 eV.}
\label{dos} 
\end{figure}

By comparison to the band stuctures of $\alpha-$ and $\beta-$As/Sb, it's found that the band gap of $\alpha-$As/Sb is direct while the band gap of $\beta-$As/Sb is indirect, and furthermore, the value of the bandgap of $\alpha-$As/Sb is smaller than that of the $\beta$ counterpart. Considering the SOC effects, it is found that the SOC effect influences the $\alpha-$ and $\beta-$Sb more significantly compared to $\alpha-$ and $\beta-$As, Which is due to the relatively heavier atoms of antimonene. 

In order to explain the formation of the band gap of $\alpha-$ and $\beta-$As/Sb and clarify contributions from different orbits, we have calculated the density of state (DOS) of monolayer $\alpha-$ and $\beta-$As/Sb, as shown in Fig.~\ref{dos}. The total and projected densities of states are calculated as shown in Fig.~\ref{dos}, which indicates that for  $\alpha-$ and $\beta-$As/Sb, the p-orbital electronic states dominate the top of the valence bands and the bottom of the conduction bands.

\begin{figure}
\centering
\includegraphics[width=0.9\linewidth]{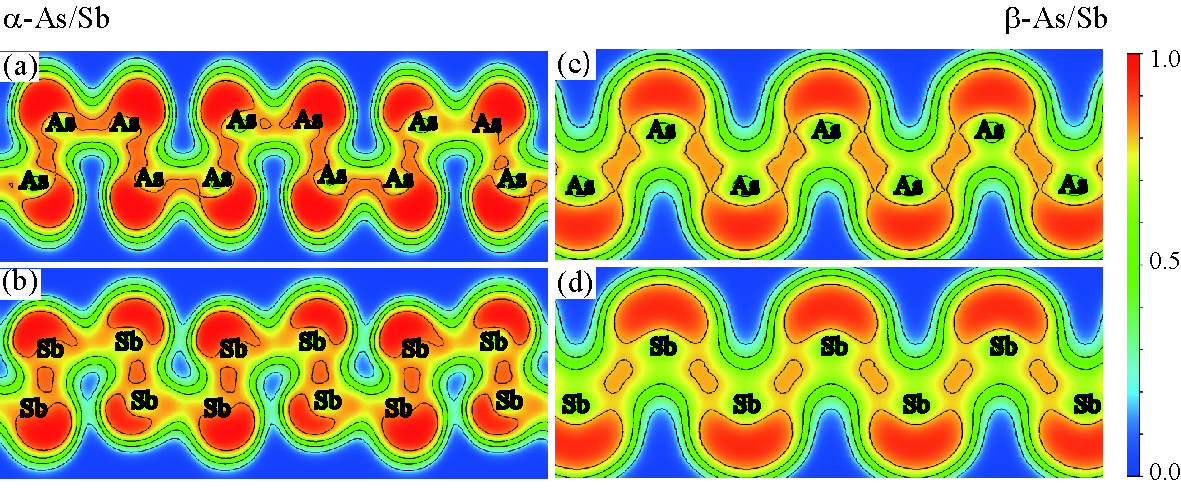}
\caption{Top view of 2D ELF profiles of monolayer (a) $\alpha-$As, (b) $\alpha-$Sb in the($\bar{1}\bar{1}\bar{1}$) plane, and (c) $\beta-$As, (d) $\beta-$Sb in the ($0\bar{1}4$) plane, respectively.}
\label{elf} 
\end{figure}

To understand the bonding characteristics, the electron localization function (ELF) \cite{Becke1990,Savin1992,Gatti2005,Chen2013} is calculated, as shown in Fig.~\ref{elf}. The ELF is a position dependent function with values that range from 0 to 1. ELF=1 corresponds to perfect localization and ELF=0.5 correponds to the electron-gas like pair probability. The respective ELF profile of $\alpha-$As/Sb, as shown in Fig.~\ref{elf}(a) and (b) respectively, is similar to that of the $\beta$ counterpart, as shown in Fig.~\ref{elf}(c) and (d) respectively. Furthermore, although the ELF values increase in regions around the As/Sb sites in all four materials, the ELF values between the atoms in different sublayers in $\alpha-$As/Sb is larger than those in $\beta-$As/Sb, which means that the chemical bondings in $\alpha-$As/Sb are more covalent compared to $\beta-$As/Sb.

\subsection{Optical properties of $\alpha-$ and $\beta-$As/Sb}

The optical properties of monolayer $\alpha-$ and $\beta-$As/Sb are described by the complex dielectric function, $i.e.$ $\epsilon(\omega)=\epsilon_1(\omega)+i\epsilon_2(\omega)$. The imaginary part of dielectric tensor $\epsilon_2(\omega)$ is determined by a summation over empty band states using as follows \cite{Gajdos2006},

\begin{equation}
\epsilon_2(\omega) = \frac{2\pi e^2}{\Omega \epsilon_0} \sum_{k,v,c} \delta(E_k^c-E_k^v-\hbar \omega) \Bigg\vert\langle \Psi_k^c \big\vert \textbf{u}\cdot\textbf{r} \big\vert \Psi_k^v \rangle \Bigg\vert ^2,
\end{equation}

where $\epsilon_0$ is the vacuum dielectric constant, $\Omega$ is the crystal volume, $v$ and $c$ represent the valence and conduction bands respectively, $\hbar\omega$ is the energy of the incident phonon, \textbf{u} is the vector defining the polarization of the incident electric field, \textbf{u}$\cdot$\textbf{r} is the momentum operator, $\Psi_k^c$ and $\Psi_k^v$ are the wave functions of the conduction and valence band at the $k$ point, respectively. The real part of dielectric tensor $\epsilon_1(\omega)$ is obtained by the well-known Kramers-Kronig relation\cite{dresselhaus1999solid},

\begin{equation}
\epsilon_1(\omega)=1+\frac{2}{\pi}P\int_0^{\infty} \frac{\epsilon_2(\omega ')\omega '}{\omega '^2-\omega^2+i\eta}d\omega ',
\end{equation}
where $P$ denotes the principle value. Based on the dielectric function of $\alpha-$ and $\beta-$As/Sb, the optical properties including the energy loss spectrum $L(\omega)$ can be subsequently given by \cite{Saha2000,Luo2015,peng2016first}

\begin{equation}
\alpha(\omega)=\frac{\sqrt{2}\omega}{c} \Big\lbrace \big[\epsilon_1^2(\omega)+\epsilon_2^2(\omega)\big]^{1/2}-\epsilon_1(\omega) \Big\rbrace ^{\frac{1}{2}},
\end{equation}
\begin{equation}
R(\omega)=\Bigg| \frac{\sqrt{\epsilon_1(\omega)+i\epsilon_2(\omega)}-1}{\sqrt{\epsilon_1(\omega)+i\epsilon_2(\omega)}+1} \Bigg| ^2,
\end{equation}
\begin{equation}
L(\omega)=Im\Big(-\frac{1}{\epsilon(\omega)}\Big)=\frac{\epsilon_2(\omega)}{\epsilon_1^2(\omega)+\epsilon_2^2(\omega)}.
\end{equation}

\begin{figure}
\centering
\includegraphics[width=0.7\linewidth]{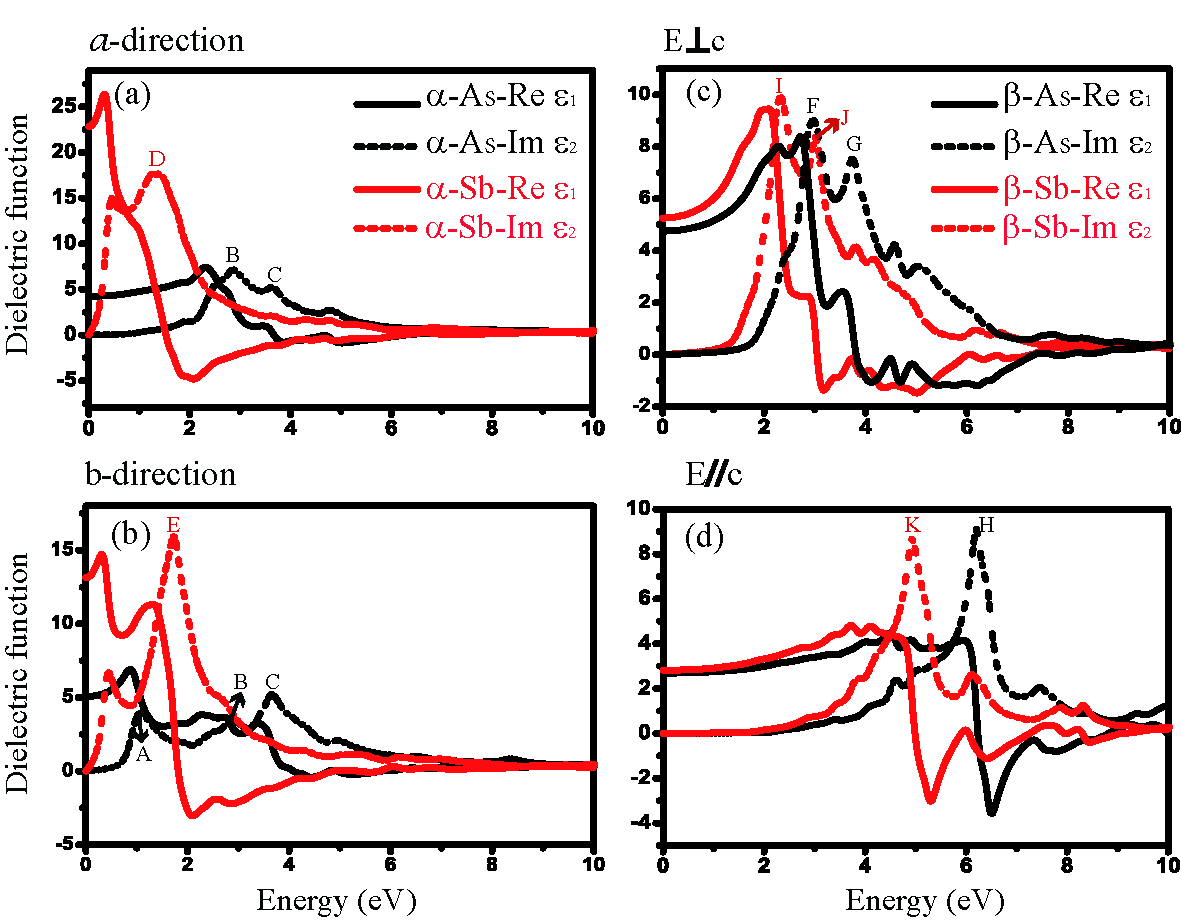}
\caption{Dielectric functions of monolayer $\alpha-$As/Sb(a,b) along the $a$ and $b$ directions, and $\beta-$As/Sb (c,d) for in-plane polarization ($\textbf{E}\perp c$) and out-plane polarization ($\textbf{E}//c$), respectively.}
\label{dielctric} 
\end{figure}

\begin{figure}
\centering
\includegraphics[width=0.7\linewidth]{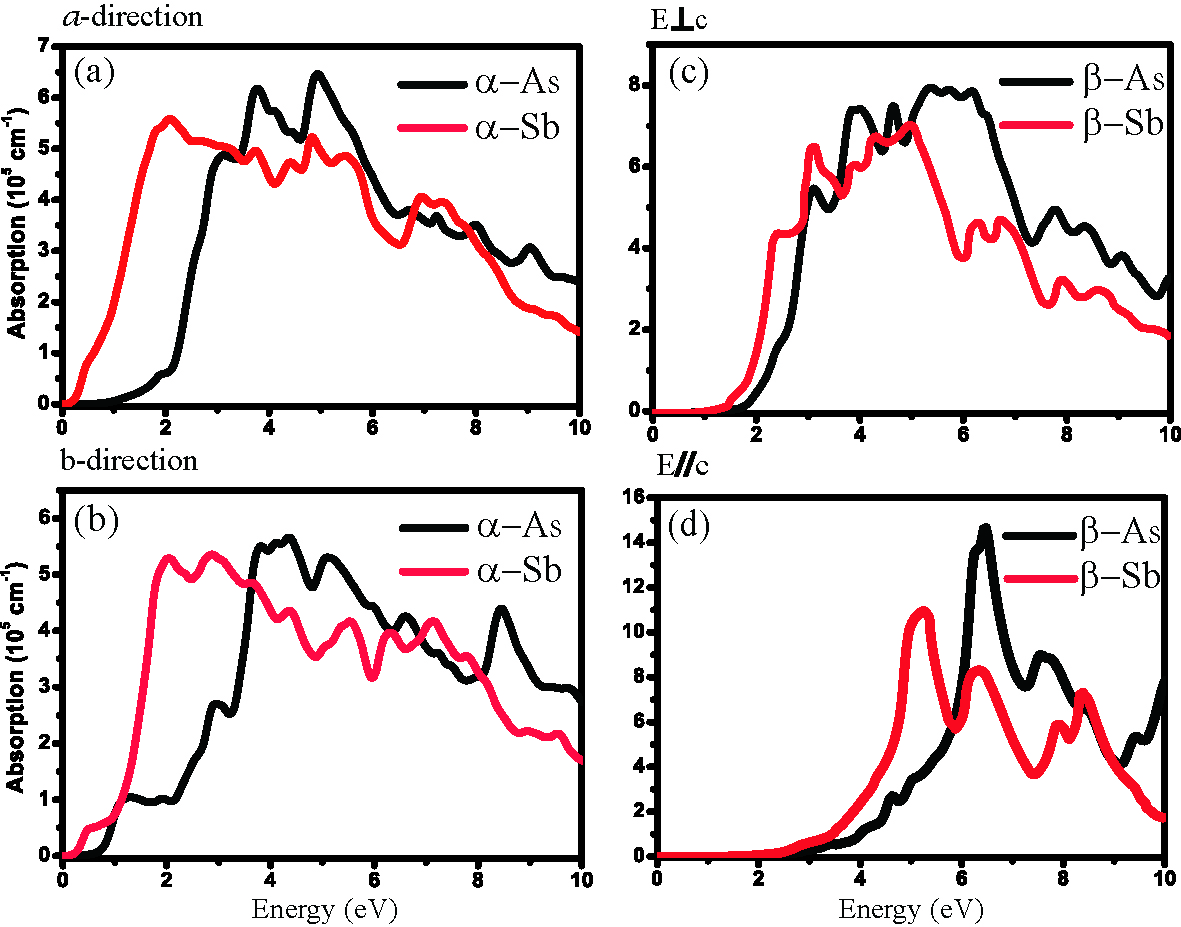}
\caption{Absorption coefficients of (a,b) $\alpha-$As/Sb along the $a$ and $b$ directions, and (c,d) $\beta-$As/Sb for $\textbf{E}\perp a$ and $\textbf{E}//a$, respectively.}
\label{absorption} 
\end{figure}

\begin{figure}
\centering
\includegraphics[width=0.7\linewidth]{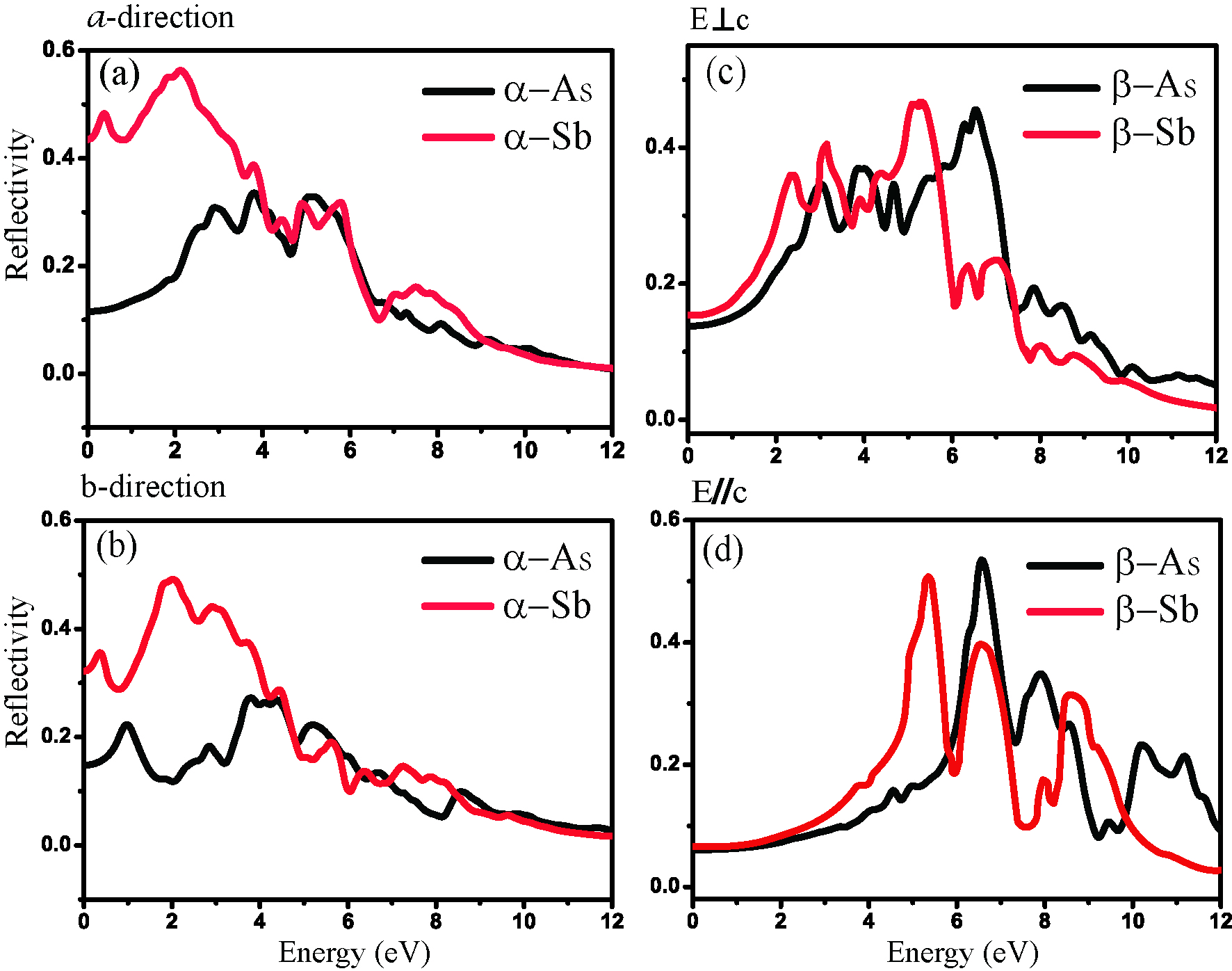}
\caption{Reflectivities of (a,b) $\alpha-$As/Sb along the $a$ and $b$ directions, and (c,d) $\beta-$As/Sb for $\textbf{E}\perp c$ and $\textbf{E}//c$, respectively.}
\label{reflectivity} 
\end{figure}

Figs.\ref{dielctric}-\ref{reflectivity} give the dielectric function $\epsilon(\omega)$, energy loss function $L(\omega)$, the absorption coefficient  $\alpha(\omega)$ and the reflectivity $R(\omega)$ are obtained for both $\alpha-$ and $\beta-$As/Sb. For the anisotropic $\alpha-$As/Sb, only in-plane optical properties are calculated for incident light with the polarization of the electric field \textbf{E} along the $a$ (\textbf{E}//$a$) and $b$ (\textbf{E}//$b$) directions. While for the isotropic $\beta-$As/Sb, optical properties are calculated for in-plane polarization ($\textbf{E}\perp c$) and out-plane polarization ($\textbf{E}//c$), respectively. 

For the dielectric function $\epsilon(\omega)$ of $\alpha-$As/Sb shown in Fig.~\ref{dielctric}(a) and (b) respectively, strong in-plane anisotropy ($\epsilon^{aa}\neq\epsilon^{bb}$) is observed, which is attributed to the anisotropic crystal structure of $\alpha-$As/Sb. As we know, the peaks in the imaginary part of dielectric function $\epsilon_2(\omega)$ are caused by the absorption of incident photons and the interband transition of electrons\cite{Mohan20121670}. 

For $\alpha-$As, the imaginary part of dielectric function $\epsilon_2(\omega)$ for \textbf{E}//$a$ shows peaks at 2.84 eV and 3.64 eV indicated by B and C respectively in Fig.~\ref{dielctric}(a), while in the curve of $\epsilon_2(\omega)$ for \textbf{E}//$b$, three main peaks at 1.0 eV, 2.86 eV and 3.66 eV indicated by A, B and C respectively are observed in Fig.~\ref{dielctric}(b). By comparison to the band structure of  $\alpha-$As shown in Fig.~\ref{band}(a), the peak A is attributed to the interband transition along $Y-X$ direction close to the $\Gamma$ point. The peak B is probably due to the ``paralle band effect" along the $X-S$ direction indicated in Fig.~\ref{band}(a) \cite{Fox2001}. The peak C is associated with the interband transition of electron from the VB to CB near Fermi energy along the $Y-X$ direction close to the $S$ point shown in Fig.~\ref{band}(a). 

For $\alpha-$Sb, in the curves of $\epsilon_2(\omega)$ shown in Fig.~\ref{dielctric}(a) and (b), for \textbf{E}//$a$, one main peak is observed at 1.33 eV (peak D); for \textbf{E}//$b$, also one main peak is observed at 1.73 eV (peak E). Both peaks are attributed to the interband transition of electron from VB to CB near Fermi energy, as shown in the band structure of $\alpha-$Sb in Fig.~\ref{band}(b).

Similar analysis can be performed to the the imaginary part of dielectric function $\epsilon_2(\omega)$ of $\beta-$As/Sb, and the peaks in the curves of  $\epsilon_2(\omega)$ of $\beta-$As/Sb, shown in Fig.~\ref{dielctric}(c) and (d) respectively, can be attributed to the corresponding interband transitions of electrons indicated in the band structures of $\beta-$As/Sb in Fig.~\ref{band}(c) and (d) respectively.

By comparison to the dielectric functions between $\alpha-$As and $\alpha-$Sb, the curves of $\epsilon_2(\omega)$ of  $\alpha-$As/Sb for \textbf{E}//$a$ and \textbf{E}//$b$, shown as the black and red dot lines in Fig.~\ref{dielctric}(a) and (b), the number of peaks and the profile of $\epsilon_2(\omega)$ curves are roughly similar, except that the $\epsilon_2(\omega)$ curves of $\alpha-$As are blueshifted to some extent to those of $\alpha-$Sb. A greater similarity and a smaller blueshift are observed between the $\epsilon_2(\omega)$ curves of $\beta-$As and  $\beta-$Sb as well, shown as the black and red dot lines in Fig.~\ref{dielctric}(c) and (d). The similarity existed within the curves of $\epsilon_2(\omega)$ results from the similarity of the band structure of $\alpha-$As ($\beta-$As) compared to that of $\alpha-$Sb ($\beta-$Sb) as mentioned above, since the imaginary part of dielectric function $\epsilon_2(\omega)$ is determined by the band structure of crystal according to Eq. (1). It is obvious that the simlarity between the band structures of $\beta-$As and $\beta-$Sb, shown in Fig.~\ref{band}(c) and (d), is greater than that of $\alpha-$As and $\alpha-$Sb, shown in Fig.~\ref{band}(a) and (b), as a result, a greater similarity between the the curves of $\epsilon_2(\omega)$ of $\beta-$As and $\beta-$Sb can be thus observed.

Since the major contribution to $\epsilon_2(\omega)$ of semiconductors comes from the interband transitions of electrons from VB to CB near Fermi energy\cite{dresselhaus1999solid}, and the band gap of $\alpha-$As is larger than that of  $\alpha-$Sb, therefore the $\epsilon_2(\omega)$ curves of $\alpha-$As is blueshifted compared to those of  $\alpha-$Sb, as shown in Fig.~\ref{dielctric}(a) and (b). The observed blueshift in the $\epsilon_2(\omega)$ curves of $\beta-$As compared to those of $\beta-$Sb, as shown in Fig.~\ref{dielctric}(c) and (d), can be explained in a similar way.   

According to the Kramers-Kronig relation, i.e. Eq. (2), the real part of the dielectric function $\epsilon_1(\omega)$ is determined by the imaginary part of the dielectric function $\epsilon_2(\omega)$, as a result, the above-mentioned similarity and blueshift can be also observed in the $\epsilon_1(\omega)$ curves of $\alpha-$As ($\beta-$As) compared to those of $\alpha-$Sb ($\beta-$Sb), shown as the black and red solid lines in Fig.~\ref{dielctric}(a-b) (Fig.~\ref{dielctric}(c-d)).

The calculated absorption coefficients for both $\alpha-$As/Sb and $\beta-$As/Sb are shown in Fig.~\ref{absorption}. The above-mentioned similarity and blueshift are observed as well in the absorption curves of $\alpha-$As ($\beta-$As) compared to those of $\alpha-$Sb ($\beta-$Sb), except that the absorption bandwidth of antimonene is broader than that of arsenene irrespective of $\alpha$ or $\beta$ structure. In addition, significant absorption from the visible region to the ultraviolet region can be observed in $\alpha-$Sb, while for $\alpha-$As the significant absorption begins in the blue region and ends in the ultraviolet region. So $\alpha-$Sb may become an alternative candidate for the application of saturable absorber which can be used in laser device.  It should be noted that, the absorption of $\beta-$As/Sb is negligible in the visible region, as shown in Fig.~\ref{absorption}(d). 

Fig.~\ref{reflectivity} shows the reflectivity for both $\alpha-$As/Sb and $\beta-$As/Sb. It is shown in Fig.~\ref{reflectivity}(a-b) that, for the case of $\alpha-$As along both $a$ and $b$ directions, the reflectivity in the visible region is high. Considering that the absorption in the visible region for $\alpha-$As is large too as mentioned above, $\alpha-$As is a non-transparent material. However, for $\beta-$As/Sb when \textbf{E}//$b$, the reflectivity in the visible region is low and the absorption in the visible region is negligible, indicating that $\beta-$As/Sb are directionally transparent materials.

\begin{figure}
\centering
\includegraphics[width=0.8\linewidth]{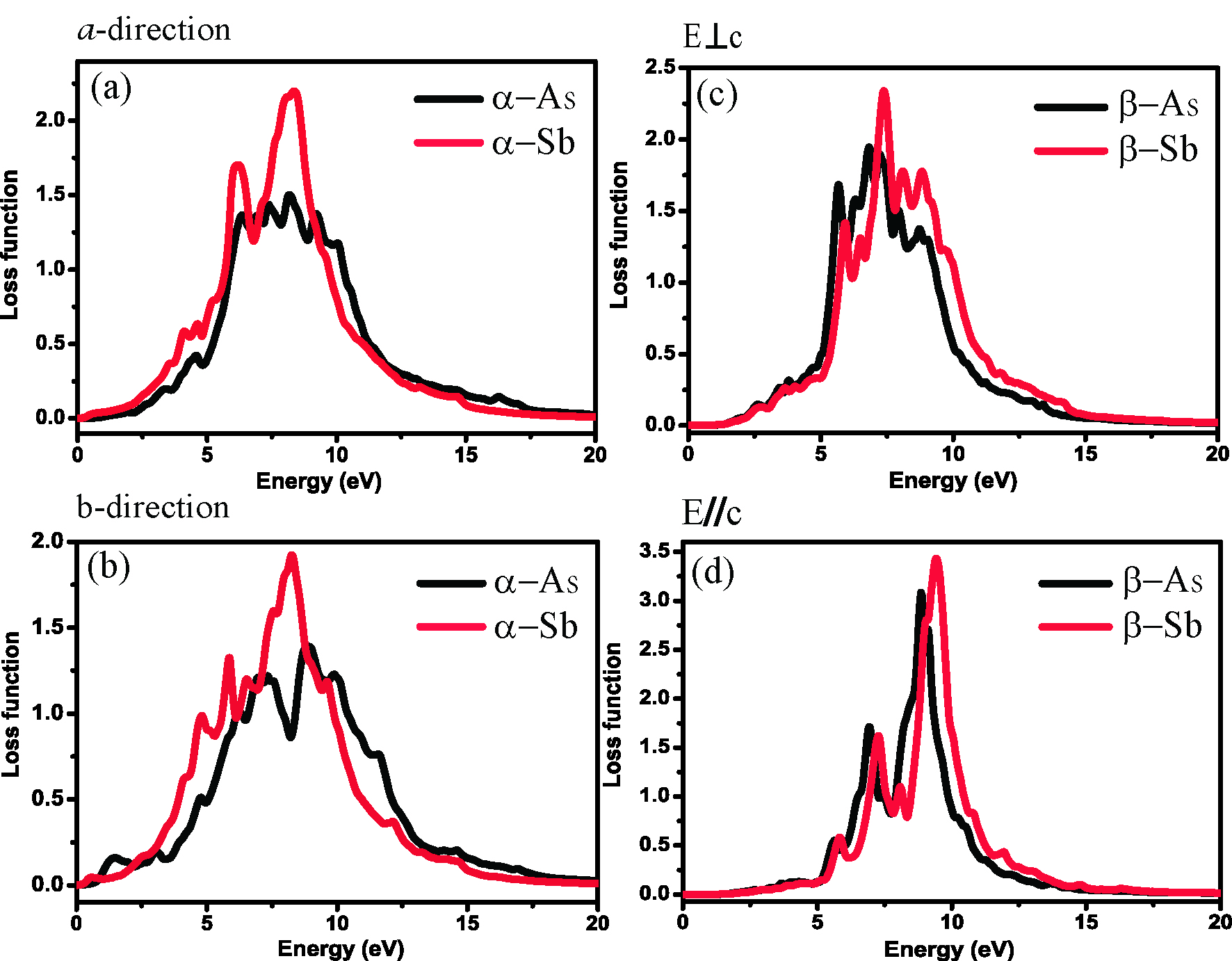}
\caption{Electron energy-loss functions of (a,b) $\alpha-$As/Sb along the $a$ and $b$ directions, and (c,d) $\beta-$As/Sb for $\textbf{E}\perp c$ and $\textbf{E}//c$, respectively.}
\label{loss} 
\end{figure}
Fig.~\ref{loss} shows the electron energy loss spectroscopy for both $\alpha-$As/Sb and $\beta-$As/Sb. Electron energy loss spectroscopy (EELS) describes the energy lost by electron when passing through dielectric materials, and can be used to deduce the dielectric function of materials\cite{C5CP00563A}.

\section{Conclusion}

In conclusion, we systematically investigate the phonon, optical and electronic properties of monolayer $\alpha-$ and $\beta-$As/Sb by first-principles calculations. The calculations on the phonon dispersion of monolayer $\alpha-$ and $\beta-$As/Sb reveal the dynamical stability of these newly proposed 2D materials. The calculated electronic band stuctures of $\alpha-$ and $\beta-$As/Sb show that, the band gap of $\alpha-$As/Sb is direct while the band gap of $\beta-$As/Sb is indirect, and furthermore, the value of the bandgap of $\alpha-$As/Sb is smaller than that of the $\beta$ counterpart. For the dielectric function $\epsilon(\omega)$ of $\alpha-$As/Sb, strong in-plane anisotropy ($\epsilon^{aa}\neq\epsilon^{bb}$) is observed, which is attributed to the anisotropic crystal structure of $\alpha-$As/Sb. The peaks in the imaginary part of dielectric function $\epsilon_2(\omega)$ of the four materials correspond well to the interband transitions of electrons. Significant absorption from the visible region to the ultraviolet region can be observed in $\alpha-$Sb which can be used as a saturable absorber, while for $\alpha-$As the significant absorption begins in the blue region and ends in the ultraviolet region. However, for $\beta-$As/Sb when the polarization direction of the incident light is along the outplane direction, the reflectivity in the visible region is high and the absorption is almost negligible in this region, indicating that $\beta-$As/Sb are directionally optically transparent materials.

\section*{Acknowledgement}
This work is supported by the National Natural Science Foundation of China under Grants No. 11374063 and 11404348, and the National Basic Research Program of China (973 Program) under Grants No. 2013CBA01505.

\end{document}